\magnification=\magstephalf
\baselineskip=12pt
\vsize=22.0truecm
\hsize=16.0truecm
\nopagenumbers
\parskip=0.2truecm
\def\vs{\vskip 0.2in}
\def\ts{\vskip 0.05in}

\def\n{\noindent}
\def\s{$\,$}
\def\prh{\s Hz$^{-1/2}$}
\def\eq{Eq.\ }
\def\eqs{Eqs.\ }
\font\bigbold=cmbx12

\ 
\vs
\centerline{{\bigbold Elimination of Clock Jitter Noise in }}

\centerline{{\bigbold Spaceborn Laser Interferometers}}
\vs
\vs
\centerline {Ronald W. Hellings}
\centerline {Jet Propulsion Laboratory, California Institute of Technology}
\centerline {Pasadena, California 91109 }

\vskip 1.5in

\n ABSTRACT:  
Space gravitational wave detectors employing laser interferometry between free-flying spacecraft differ in many ways from their laboratory counterparts.  Among these differences is the fact that, in space, the end-masses will be moving relative to each other.  This creates a problem by inducing a Doppler shift between the incoming and outgoing frequencies.  The resulting beat frequency is so high that its phase cannot be read to sufficient accuracy when referenced to state-of-the-art space-qualified clocks.  This is the problem that is addressed in this paper.  We introduce a set of time-domain algorithms in which the effects of clock jitter are exactly canceled.  The method employs the two-color laser approach that has been previously proposed, but avoids the singularities that arise in the previous frequency-domain algorithms.  In addition, several practical aspects of the laser and clock noise cancellation schemes are addressed.

\vfill
\eject

\centerline{{\bf I.  Background}}

\vs
\n {\bf A.  Introduction}
\ts
\nobreak
During the last ten years, work has been going on to define and design a spaceborne laser interferometer for the purposes of the detection of low frequency ($10^{-4}{\rm Hz}\rightarrow 1 {\rm Hz}$) gravitational waves [1-3].  These interferometers work by passing laser signals between widely separated spacecraft, reading out the relative phases of the signals, and combining signals between different spacecraft to eliminate laser phase noise and enhance the gravitational wave signal.  A typical design concept is shown in Figure 1.  Three spacecraft (S/C) move on trajectories that keep them at the vertices of an equilateral triangle, and signals are passed in both directions along each of the three long arms thus formed.  The passage of a gravitational wave through the system will create small changes in the curvature of space through which the laser signals are passing, thereby advancing or retarding the phases of the laser signals.  The amplitude of the phase shift produced is proportional to the amplitude of the gravitational wave, proportional to the distance between the spacecraft, and dependent on the orientation of the arm relative to the direction of propagation of the wave.  The detection of the phase differences between the signals in the arms constitutes the detection of the gravitational wave, and the detailed waveform observed for the phase change reveals information about the astronomical source that created the wave -- thus providing gravitational astronomy observations of the accelerated massive bodies that emit gravitational waves.

The difficulty in gravitational wave detection is the smallness of the phase change expected from the reasonable astronomical sources one might expect to see.  In one mission concept, the Laser Interferometer Space Antenna (LISA), the lengths of the arms are $5\times10^9\ $m.  Despite this long baseline, the amplitudes of the expected sources are so small that a sensitivity\footnote{$^1$}
  {It has long been the practice of the 
  gravitational wave community to characterize noise or 
  sensitivity in terms of the ``root spectral density''.  
  The relation between the variance and the spectral 
  density of a time series $n$ is given by 
  $\langle n^2\rangle=\int_{\Delta f}S_n(f) df$, where $\Delta f$
  is the bandwidth and $S_n$ is the spectral density with units 
  (n-units)$^2$Hz$^{-1}$.  The root spectral density is simply 
  $\sqrt{S_n(f)}$ and has units (n-units)\prh.  Over a 
  bandwidth where the spectrum is flat, the relationship between 
  the rms amplitude of $n$ and the root spectral density of $n$ is 
  $n_{\rm rms}=\sqrt{S_n(f)}\sqrt{\Delta f}$.}
of 10\s pm\prh, or 10 $\mu$cycle\prh of a 1\s $\mu$m wavelength laser, has been set as a requirement for the detector.  Unfortunately, the most stable lasers that can be built have phase fluctuations many orders of magnitude larger than this.  In ground-based gravitational wave detectors, where the phase requirements are more stringent still, the problem of laser phase fluctuations has been solved by creating an equal-arm interferometer.  In this type of instrument, a single laser signal goes down two arms, bounces off the end masses, and is recombined by allowing the returning signals to interfere with each other.  By maintaining the armlengths strictly equal, the laser phase noise cancels when the returned signals are combined.  To see how this happens, we consider two arms radiating from S/C 1 in Fig. 1.  Let the phase of the laser in S/C 1 be $\phi_1(t)$ and the gravitational wave signal in the arm of the detector between S/C 1 and S/C 2 be $h_{12}(t)$.  Then the signal received at S/C 1, assuming that the signal is sent from 1, transponded with no change of phase at S/C 2, and then beat again against the laser in 1, is
$$ y_{21}(t)=\phi_1(t-2L_{12})-\phi_1(t)+h_{12}(t)+n_{21}(t) \eqno(1) $$
where $n_{21}(t)$ is the total non-laser phase noise in S/C 1 when tracking S/C 2 and where the light travel time in the arm is $L_{12}$.  The signal received from S/C 3 is written by letting $2\rightarrow3$ in \eq 1.  When the signals received at S/C 1 from the spacecraft at the ends of the two arms are subtracted, the resulting interferometer signal is
$$z_1(t)\equiv y_{21}(t)-y_{31}(t)=\phi_1(t-2L_{12})-\phi_1(t-2L_{13})
+h_{12}(t)-h_{13}(t)+n_{21}(t)-n_{31}(t) \eqno(2)$$
If the two arms are exactly the same length ($L_{12}=L_{13}$), the laser phase noise will disappear.  However, when the ends of the arms are free-flying spacecraft, the armlengths cannot be controlled to maintain $L_{12}= L_{13}$.  In this case another method must be found.

\vs
\n {\bf B.  Unequal-Arm Interferometer Algorithms}
\ts
\nobreak
Such a method was discovered by Jim Faller and refined and published in 1996 [4].  In this method it was recognized that, since the laser phase noise in $\phi_1(t)$ is many orders of magnitude larger than any of the other noise sources, the signal in one arm of the interferometer (\eq 1) can be used to determine the laser phase noise, and the measured time series of the noise can then be used to correct the interferometer signal (\eq 2) for the fact that the arm lengths are not equal.  Working in the frequency domain, one writes the connection between $\phi_1$ and $y_{21}$ as
$$y_{21}(f)=\left(1-e^{\displaystyle -2\pi ifL_{12}}\right)\phi_1(f),\eqno(3)$$
where the Fourier decompositions are given by
$$y_{21}(f)=\int_0^\infty y_{21}(t)e^{\displaystyle 2\pi ift}dt 
 \quad\quad {\rm and} \quad \quad 
\phi_1(f)=\int_0^\infty \phi_1(t)e^{\displaystyle 2\pi ift}dt \eqno(4)$$
The Fourier composition of the laser phase noise is then found by dividing the Fourier decomposition of the observed $y_{21}$ by the transfer function, 
$$ \phi_1(f)={y_{21}(f)\over1-e^{\displaystyle -2\pi ifL_{12}}}  \eqno(5)$$
and the time series for $\phi_1(t)$ is generated by Fourier synthesis from $\phi_1(f)$.  Once $\phi_1(t)$ is known, its contribution to \eq 2 can be synthesized and subtracted from the observed $z_1(t)$ to give an interferometer signal that is free of laser phase noise.

The problem in this procedure, however, is that \eq 5 has singularities at frequencies $f = n/L$, where $n$ is a positive integer,\footnote{$^2$}
  {\eq 5 is also singular at $f=0$ (when $n=0$), 
  but the laser phase noise in the interferometer signal (\eq 2)
  goes to zero as $f\rightarrow0$, so the detector sensitivity 
  remains unimpaired in the low frequency limit.}
so the gravitational wave detector will be relatively insensitive near these frequencies.  In addition, this method requires a Fourier transform, with all its sensitivity to biases and aliasing, before the laser phase correction can be implemented.  Fortunately, a new method has recently been discovered that works entirely in the time domain [5] and avoids these difficulties.  This method consists of combining the signals from each arm with a time-offset in such a way as to undo the inherent time-offsets at the light-times ($2L_{ij}$) in \eq 2.  The combination is named $X(t)$ and is given by
$$ X(t)=y_{21}(t-2L_{13})-y_{31}(t-2L_{12})-y_{21}(t)+y_{31}(t) \eqno(6)$$
A little algebra will show that the combination of signals in \eq 6, using \eq 1 for $y_{ij}(t)$, will exactly eliminate the $\phi_1(t)$ terms.  The $y_{ij}$ in \eq 6 represent round-trip signals.  In a recent paper, Armstrong, Estabrook and Tinto [6], have used one-way signals and identified several more combinations of signals that will likewise cancel out the $\phi_i(t)$ noise terms without cancelling the gravitational wave signals.

\vs
\n {\bf C.  High Doppler Rate Algorithms}
\ts
\nobreak
As a result of these procedures, laser phase noise can be essentially eliminated as a noise source.  However, there remains another noise source, whose level is likewise many orders of magnitude greater than the desired noise floor, that must be addressed.  This noise arises due to the fact that the relative velocity of free-flying spacecraft will produce a Doppler shift in the frequency of the signal received at each spacecraft, so that the beat frequency between the received laser signal and the local laser signal will amount to tens of MHz.  In addition, if the lasers in the two spacecraft are each independently stabilized by their own Fabry-Perot cavities, then the frequencies of the two lasers, being determined by the lengths of the cavities which cannot be made exactly equal, will differ by even more, probably by several hundred MHz.  The gravitational wave will appear as a tiny ($\sim10\, \mu{\rm cycle}$\prh) shift in the phase of this radio-frequency (RF) beat signal.  As we discuss in the next section, the measurement of phase to this precision in a signal at this high a frequency will require a frequency standard with relative frequency stability $\Delta\nu/\nu=10^{-17}$ over a time equal to the period of the gravitational wave we are trying to detect ($\sim1000\,$s).  This is beyond the capability of the best laboratory frequency standards, or `clocks', and well beyond the capability of those that can reasonably be used in space.  So what is to be done?

In a companion paper [7] to the frequency-domain unequal-arm algorithm paper [4], a frequency-domain procedure was presented that would eliminate this clock noise.  This method required that a second laser signal be used along the arms of the interferometer.  Each spacecraft must therefore broadcast two laser signals, the main signal and a second signal that is frequency-offset relative to the main signal by an amount tied to the local clock.  Thus, whatever phase reference is being used at one spacecraft is sent off to the far spacecraft where it is measured and recorded.  Then, working in the frequency domain, an inversion like that in \eq 5 may be performed, and a time series giving the jitter in each clock may be formed by Fourier synthesis.  In this way, whatever error is induced by reading laser phase relative to a noisy frequency standard may be simulated and corrected.  This frequency-domain method, however, suffers all of the limitations noted above for synthesizing $\phi_i(t)$, including the poles at $f=n/L$.

It is the main purpose of this paper to present a set of time-domain algorithms that may be used to process the dual-frequency laser data and eliminate the noise due to clock jitter.  These algorithms will be developed in Section II and will represent the clock-jitter counterparts of the laser phase noise algorithms reported in references [5] and [6].  In Section III, we also present several practical aspects of the data processing that have not previously been addressed.

\vfill
\eject

\centerline{{\bf II.  Time-Domain Algorithms}}

\vs
\n {\bf A.  Instrument Concept}
\ts
\nobreak
We assume an instrument labeled as in Fig. 1, with three identical spacecraft numbered 1, 2, and 3.  The signal received at time $t$ by S/C 1 from S/C 2 will be denoted as $y_{21}(t)$.  In the previous section, our expressions for the signals assumed that the lasers in S/C 2 and S/C 3 were phase-locked to the incoming signals from S/C 1, so that whatever phase S/C 2 received was simply bounced back toward S/C 1.  While this assumption simplifies the formulas in the case of a single interferometer with a single central spacecraft, there are reasons [6,8] for keeping the lasers in each spacecraft independent of the others.  Indeed, if data from a second interferometer (with vertex at one of the other spacecraft) are to be collected simultaneously with the first, then the forms of the signals for the second interferometer are not the simple expressions given in Section I.  In the remainder of this paper we will assume that the laser in each spacecraft is locked only to its own Fabry-Perot cavity.  The case where some lasers are locked to incoming signals may be recovered as a special case of the expressions we derive (see Section III.D).

We write the phase of the signal sent by S/C $i$ and received by S/C $j$ as
$$ y_{ij}(t)=\phi_i(t-L_{ij})-\phi_j(t)+h_{ij}(t)+n_{ij}(t) \eqno(7) $$
The notation is the same as in \eq 1, except that $h_{ij}$ and $n_{ij}$ have changed their meaning slightly.  In \eq 1, these quantities were the total signal and the total non-laser noise generated during the round-trip of the signals, including whatever noise S/C $i$ contributed while it was transponding the signal.  Here, in \eq 7, $h_{ij}$ represents the gravitational wave signal generated during the one-way trip from S/C $i$ to $j$, and $n_{ij}$ represents the noise that arises one-way, dominated at low Fourier frequencies by equal amounts of position noise in both spacecraft and at high Fourier frequencies by shot noise in the laser receiver at S/C $j$ [9].  We further separate the laser phase into a part that is a pure constant frequency and a part that is random phase noise
$$ \phi_i(t)=\nu_it+p_i(t) \eqno(8) $$
where $\nu_i$ is the constant laser frequency in S/C $i$ and $p_i(t)$ is the phase noise in this laser.  If we also write the armlength as an initial distance plus a constant velocity,
$$ L_{ij}=L_{ij,0}+V_{ij}t \eqno (9)$$
then \eq 7, with \eqs 8 and 9 included, becomes
$$ y_{ij}(t)=\left[(\nu_i-\nu_j-V_{ij}\nu_i\right]t
+p_i(t-L_{ij})-p_j(t)+h_{ij}(t)+n_{ij}(t) \eqno(10) $$
where we have dropped the constant phase offset associated with $L_{ij,0}$.  

Now let us estimate the sizes of the various terms in \eq 10.  The magnitude of the first term, $\nu_i-\nu_j$, will depend on how closely tuned the frequencies of the lasers in the two spacecraft may be, considering that each will be locked to its own Fabry-Perot cavity.  For the OMEGA mission [3], which proposed using independent Nd:YAG (1064\s $\mu$m wavelength) lasers in each spacecraft in this way, a reasonable estimate was found to be 300 MHz.  In any event, the next term, $V_{ij}\nu_i$, will be of order $\sim10$\s MHz, so the receiver design must accomodate this fundamental frequency for $y_{ij}$, even in the case of non-independent end lasers.  The size of the $p_i$ terms in \eq 10 will depend on how accurately the lasers are locked to the cavities in each spacecraft (see Section III.A for an estimate).  The smallest detectable $h_{ij}(t)$ gravitational wave phase signal is required to be 10\s$\mu$cycles\prh, so uncancelled $p_i(t)$ laser noise and other $n_{ij}(t)$ phase noise must be less than this.  Therefore, in order to detect a gravitational wave with period $\tau=1000\,$s in a bandwidth $\Delta f=1/\tau$, a phase shift of $\delta\phi=\sqrt{S_\phi}\sqrt{\Delta f}=0.3\,\mu$cycle must be detected in a total phase of $300\,{\rm MHz}\times1000\,{\rm s}=3\times10^{11}\,$cycles.  If we recast this requirement in terms of frequency, we see that this corresponds\footnote{$^3$}
  {The spectral density of frequency noise is related to 
  the spectral density of phase noise by 
  $S_\nu=\omega^2S_\phi=4\pi^2f^2S_\phi$, 
  so the rms frequency noise is
  $\delta\nu_{\rm rms}=\sqrt{S_\nu}\sqrt{\Delta f}
  = 2\pi f\,\delta\phi_{\rm rms}$.}  
to detecting a 2\s nHz frequency shift in a signal of 300\s MHz.  In order to measure a frequency to a this accuracy, a frequency standard with frequency stability $2\,{\rm nHz}/300\,{\rm MHz}\approx10^{-17}$ is required.  Since there presently exists no space-qualified frequency standard with such performance, clock jitter will be a major noise source in the detectors unless it is dealt with in some other way.

\vs
\n {\bf B.  Clock Jitter}
\ts
\nobreak
To see how clock jitter noise enters the readout of the laser phases, let us consider the detection process in more detail.  We will actually so far anticipate our final solution to this problem as to discuss the two laser signals that will ultimately be needed.  Figure 2 shows symbolically the signals that are sent and received by the two spacecraft at the two ends of a single arm.  The fundamental laser frequency in S/C 2 is $\nu_2$ and the laser phase noise is $p_2(t)$.  In addition, a second laser signal is superimposed on the same beam, either by merging signals from two separate lasers (as in OMEGA [3], with the beat frequency between the two acting as the local spacecraft clock) or by modulating the main laser signal at a frequency equal to the fundamental frequency of the ultra-stable oscillator (USO) that serves as the local spacecraft clock (as in the current LISA design [2]).  The frequency of the local clock is $f_2$ and it is assumed to have phase jitter $q_2(t)$.  Thus S/C 2 will send laser signals $\nu_2t+p_2(t)$ and $(\nu_2+f_2)t+p_2(t)+q_2(t)$, while S/C 1 has local laser signals $\nu_1t+p_1(t)$ and $(\nu_1+f_1)t+p_1(t)+q_1(t)$.  The phases of the two signals from S/C 2, as received at S/C 1, are
$$ \phi_2(t)=\nu_2(1-V_{12})t+p_2(t-L_{12})+h_{21}(t)+n_{21}(t) \eqno(11a) $$
and
$$ \phi'_2(t)=(\nu_2+f_2)(1-V_{12})t+p_2(t-L_{12})+q_2(t-L_{12})
  +h_{21}(t)+n'_{21}(t) \eqno(11b) $$
The $n'_{21}$ in \eq 11b may include some noise, like spacecraft position noise, that is the same as in $n_{21}$, and other noise, like shot noise, that will be different from the shot noise part of $n_{21}$.  When these signals are beat against the local laser at S/C 1, there result the two signals 
$$ y_{21}(t)=\phi_2(t)-\phi_1(t)=\left[(\nu_2-\nu_1)-V_{12}\nu_2\right]t
+p_2(t-L_{12})-p_1(t)+h_{21}(t)+n_{21}(t) \eqno(12a) $$
and
$$ \eqalign{
y'_{21}(t)=\phi'_2(t)-\phi'_1(t)=&
\left[(\nu_2-\nu_1)+(f_2-f_1)-V_{12}(\nu_2+f_2)\right]t \cr
&+p_2(t-L_{12})-p_1(t)+q_2(t-L_{12})-q_1(t)+h_{21}(t)+n'_{21}(t)} \eqno(12b) $$
both composed of constant-frequency terms plus phase fluctuations.

The procedure for reading out the phases of the two signals $y_{21}(t)$ and $y'_{21}(t)$ may best be understood by reference to Fig. 3.  The two incoming signals from S/C 2, $\phi_2$ and $\phi'_2$, fall on a photodiode on board S/C 1, where they interfere with a portion of the two signals, $\phi_1$ and $\phi'_1$, that are being broadcast from S/C 1.  If, as has been proposed [7], the RF clock signals $f_i$ are at much higher frequency than the 300 MHz of $\nu_2-\nu_1$, then $y_{21}(t)$ and $y'_{21}(t)$ will be the only signals in this frequency band (see reference [7], \eq 14) and will not be confused with any other of the six beat frequencies on the photodiode, generated by mixtures of the four input frequencies.

In order to read out these RF frequencies, a local oscillator (LO) is first used to beat the signals down to a low baseband where they may be sampled and fit to determine an average phase.  For the clock jitter cancellation procedure to work, the LO must be phase-locked to the local $f_i$ clock, as we show below.  The LO could be a piece of hardware (like the phase-stable frequency synthesizer developed for the OMEGA mission [10]) or a piece of software in a digital processor (as long as the clock cycles for the processor are tied to the spacecraft time standard).  Whatever the realization of this frequency subtractor, however, it is important for the method of this paper that the same frequency be subtracted from both $y_{21}(t)$ and $y'_{21}(t)$.\footnote{$^4$}
  {An example of a receiver that does not 
  satisfy this requirement would be one that processed 
  the RF signals directly to measure their phase.  Such 
  an instrument can be thought of as an LO that beats each 
  signal separately down to DC.  In order to do this for both 
  $y_{21}(t)$ and $y'_{21}(t)$, a different frequency would 
  have to be mixed with $y_{21}(t)$ than with $y'_{21}(t)$, 
  and the clock jitter cancellation procedure we describe here 
  will not work at the required accuracy.}

The LO will produce a frequency that is close to the frequencies of the two signals $y_{21}(t)$ and $y'_{21}(t)$.  This frequency will be tied to the local spacecraft clock by making the LO frequency some rational fraction $a_{21}$ of the local clock frequency $f_1$.  The functional forms of the two baseband signals coming out of the mixer will then be
$$\eqalign{
s_{21}(t)&=y_{21}(t)-a_{21}\left[f_1t+q_1(t)\right]\cr
         &=\left[(\nu_2-\nu_1)-V_{12}\nu_2-a_{21}f_1\right]t
+p_2(t-L_{12})-p_1(t)-a_{21}q_1(t)+h_{21}(t)+n_{21}(t)
}\eqno(13a)$$
and
$$\eqalign{
s'_{21}(t)&=y'_{21}(t)-a_{21}\left[f_1t+q_1(t)\right]\cr
         &=\left[(\nu_2-\nu_1)+(f_2-f_1)-V_{12}(\nu_2+f_2)-a_{21}f_1\right]t\cr
         &\quad\quad\quad\quad+p_2(t-L_{12})-p_1(t)+q_2(t-L_{12})
                 -(a_{21}+1)q_1(t)+h_{21}(t)+n'_{21}(t)
}\eqno(13b) $$
where $a_{21}$ has been chosen so that $a_{21}f_1\sim\nu_2-\nu_1-V_{12}\nu_2$.  \eq 13a is the main signal, which will be present even if there is no second laser frequency in the beam.  The clock jitter problem we are trying to solve is apparent in this equation.  When any laser signal is detected and read out, the $q_1(t)$ jitter in the local clock in the receiving spacecraft will produce a phase noise $a_{21}q_1(t)$ in the result.  Assuming a standard space-qualified USO, this term will correspond to frequency noise $\sim1\,\mu$Hz.  This is three orders of magnitude larger than the 2\s nHz accuracy needed for the gravitational wave requirement.  Therefore, even after the $X(t)$ combination is formed and the $p_i(t)$ terms drop out, there will be $a_{ij}q_i(t)$ terms that will remain and will dominate the noise in the detector.  

\vs
\goodbreak
\n {\bf C.  Laser and Clock Noise Cancellation Algorithms}
\ts
\nobreak
Before we discuss the algorithms that will cancel the $q_i(t)$ clock jitter noise terms in \eq 13, let us review the combinations of the $y_{ij}(t)$ signals that are used to cancel laser phase noise.  We summarize the results from the papers by Armstrong, Estabrook and Tinto [5,6], but we use notation (for numbering spacecraft and arms) consistent with Fig. 1 and \eq 7.

Consider, then, the combination of $y_{ij}(t)$ defined by
$$\eqalign{
X(t)=y_{12}&(t-L_{12}-2L_{13})-y_{13}(t-L_{13}-2L_{12})+y_{21}(t-2L_{13})\cr
      &-y_{31}(t-2L_{12})+y_{13}(t-L_{13})-y_{12}(t-L_{12})+y_{31}(t)-y_{21}(t)
}\eqno (14a) $$
It may be noted that $X(t)$ is formed from signals that travel only along the two arms $L_{12}$ and $L_{13}$, with no data included from the $L_{23}$ arm.  If there were ever a failure in the spacecraft such that one arm of the interferometer would be no longer operative, that arm could be designated the $L_{23}$ arm, and the other two arms could still be used to provide a laser-phase-noise-free $X(t)$ signal.  Functions analogous to $X(t)$ may also be formed by permuting each subscript in \eq 14 ($1\rightarrow2\rightarrow3$) to define a signal $Y(t)$ which requires no data from $L_{13}$, 
$$\eqalign{
Y(t)=y_{23}&(t-L_{23}-2L_{12})-y_{21}(t-L_{12}-2L_{23})+y_{32}(t-2L_{12})\cr
      &-y_{12}(t-2L_{23})+y_{21}(t-L_{12})-y_{23}(t-L_{23})+y_{12}(t)-y_{32}(t)
}\eqno (14b) $$
and again to give $Z(t)$ which requires nothing from $L_{12}$.
$$\eqalign{
Z(t)=y_{31}&(t-L_{13}-2L_{23})-y_{32}(t-L_{23}-2L_{13})+y_{13}(t-2L_{23})\cr
      &-y_{23}(t-2L_{13})+y_{32}(t-L_{23})-y_{31}(t-L_{13})+y_{23}(t)-y_{13}(t)
}\eqno (14c) $$
To see how the these combinations work to eliminate laser phase noise, we expand \eq 14a, using \eq 7 to express the $y_{ij}(t)$ in terms of their elements, and find
$$\eqalign{
X(t)&=\phi_1(t-2L_{12}-2L_{13})-\phi_2(t-L_{12}-2L_{13})
          +h_{12}(t-L_{12}-2L_{13})+n_{12}(t-L_{12}-2L_{13})\cr
    &-\phi_1(t-2L_{13}-2L_{12})+\phi_3(t-L_{13}-2L_{12})
          -h_{13}(t-L_{13}-2L_{12})-n_{13}(t-L_{13}-2L_{12})\cr
    &+\phi_2(t-2L_{13}-L_{12})-\phi_1(t-2L_{13})
          +h_{12}(t-2L_{13})+n_{21}(t-2L_{13})\cr
    &-\phi_3(t-2L_{12}-L_{13})+\phi_1(t-2L_{12})
          -h_{13}(t-2L_{12})-n_{31}(t-2L_{12})\cr
    &+\phi_1(t-2L_{13})-\phi_3(t-L_{13})
          +h_{13}(t-L_{13})+n_{13}(t-L_{13})\cr
    &-\phi_1(t-2L_{12})+\phi_2(t-L_{12})
          -h_{12}(t-L_{12})-n_{12}(t-L_{12})\cr
    &+\phi_3(t-L_{13})-\phi_1(t)
          +h_{13}(t)+n_{31}(t)\cr
    &-\phi_2(t-L_{12})+\phi_1(t)
          -h_{12}(t)-n_{21}(t)
}\eqno(15) $$
As may be seen by inspecting the first two terms on each line of \eq 13, the laser phase noise terms cancel in pairs.  Similar expansions may be used to see the cancellation in $Y(t)$ and $Z(t)$ by permuting $1\rightarrow2\rightarrow3$ in \eq 15.

However, as we saw in the last section, it is not the $y_{ij}(t)$ signals that are directly measured on each spacecraft, but, rather, the $s_{ij}(t)$ signals which contain clock jitter noise in addition to the other noise terms.  Because one must use the observable $s_{ij}(t)$ to form $X(t)$, there will be unavoidable clock jitter noise included from each of the spacecraft.  Thus, the creation of
$$ \eqalign{
X(t)=s_{12}&(t-L_{12}-2L_{13})-s_{13}(t-L_{13}-2L_{12})+s_{21}(t-2L_{13})\cr
&-s_{31}(t-2L_{12})+s_{13}(t-L_{13})-s_{12}(t-L_{12})+s_{31}(t)-s_{21}(t)
} \eqno (16)$$
yields a signal with clock jitter
$$ \eqalign{
X(t)=a_{12}\left[q_2(t-L_{12})-q_2(t-L_{12}-2L_{13})\right]
      -a_{13}\left[q_2(t-L_{13})-q_3(t-L_{13}-2L_{12})\right]\cr
     +a_{21}\left[q_1(t)-q_1(t-2L_{13})\right]
     -a_{31}\left[q_1(t)-q_1(t-2L_{12})\right]+ {\rm signal} + {\rm noise}
}\eqno(17)  $$
where the signal and noise terms are the combinations of the $h_{ij}(t)$ and $n_{ij}(t)$ terms given in \eq 15, and where terms representing constant phase shifts have been dropped.  The permutations $1\rightarrow2\rightarrow3$ in \eq 17 give the residual clock jitter noise terms in $Y(t)$ and $Z(t)$.

The key to the elimination of the $a_{ij}q_k$ noise terms in \eq 17 is found by reference to \eqs 13.  When the $s'_{ij}(t)$ secondary laser signal (\eq 13b) is subtracted from the main laser signal $s_{ij}(t)$ (\eq 13a) the resulting difference\footnote{$^5$}
  {It is here that the need for a common LO frequency, $a_{ij}f_j$,
  mixed with both the $y_{ij}$ and the $y'_{ij}$ signals becomes 
  apparent, for only then will the $a_{ij}q_j$ terms cancel exactly, 
  leaving the simple combination of $q_j$ given in \eq 18.}
$$ r'_{ij}(t)\equiv s_{ij}(t)-s'_{ij}(t)=
(f_i-f_j+V_{ij}f_j)t+q_j(t)-q_i(t-L_{ij})+n_{ij}(t)+n'_{ij}(t) \eqno(18)  $$
will contain a combination of the two clock jitters.  The clock frequencies $f_i$ will be known {\it a priori} and the orbits will be known well enough to determine $V_{ij}$.  The constant frequency part of \eq 18 may therefore be subtracted off by hand, leaving a signal that contains only the $q_i$ terms plus the instrumental noise terms:
$$ r_{ij}(t)\equiv r'_{ij}(t)-(f_i-f_j+V_{ij}f_j)t=
   q_j(t)-q_i(t-L_{ij})+n_{ij}(t)+n'_{ij}(t) 
\eqno(19)$$
A little algebra will then verify that the combination
$$ \eqalign{
\xi(t)=X&(t)-a_{12}r_{21}(t-2L_{13})+a_{13}r_{31}(t-2L_{12})
     -(a_{12}+a_{21})r_{13}(t-L_{13})\cr
&+(a_{13}+a_{31})r_{12}(t-L_{12})+(a_{12}+a_{13}+a_{31})r_{21}(t)
     -(a_{13}+a_{12}+a_{21})r_{31}(t)
}\eqno(20a) $$
will exactly cancel out the $q_i$ terms in \eq 17.  Similarly, combinations given by
$$ \eqalign{
\psi(t)=Y&(t)-a_{23}r_{32}(t-2L_{12})+a_{21}r_{12}(t-2L_{23})
     -(a_{23}+a_{32})r_{21}(t-L_{12})\cr
&+(a_{12}+a_{21})r_{23}(t-L_{23})+(a_{23}+a_{12}+a_{21})r_{32}(t)
     -(a_{21}+a_{23}+a_{32})r_{12}(t)
}\eqno(20b) $$
and
$$ \eqalign{
\zeta(t)=Z&(t)-a_{31}r_{13}(t-2L_{23})+a_{32}r_{23}(t-2L_{13})
     -(a_{13}+a_{31})r_{32}(t-L_{23})\cr
&+(a_{23}+a_{32})r_{31}(t-L_{13})+(a_{31}+a_{23}+a_{32})r_{13}(t)
     -(a_{32}+a_{13}+a_{31})r_{23}(t)
}\eqno(20c) $$
will produce signals based on $Y(t)$ and $Z(t)$ that are cleaned of clock jitter noise.  It is also apparent that, if the $a_{ij}$ coefficients are much less than unity, the additional noise ($n_{ij}$ and $n'_{ij}$) contributed by adding in the $r_{ij}$ signals will be negligible compared to that already present in the signals $X(t)$, $Y(t)$, and $Z(t)$.  In fact, the $s'_{ij}$ signals only appear multiplied by $a_{ij}$, so, with $a_{ij}$ small enough, their phase noise $n'_{ij}$ can actually be much larger than $n_{ij}$ without adding appreciably to the noise in the final signals.

In order for the phase combinations of \eq 20 to be calculated, the ratios $a_{ij}$ will have to be known.  If these ratios are determined via phase-locked loops on board each spacecraft, then those values will have to be telemetered from each spacecraft so that the combinations in \eqs 20 can be formed.  However, since the LO on each spacecraft is only required to beat the frequency down to a baseband of, perhaps, a few kHz, accurate enough values for $a_{ij}$ may easily be determined on the ground from knowledge of the clock frequencies and of the spacecraft orbit.  In this case the proper ratios, including Doppler shifts, may simply be uploaded from the ground and stored in memory on board each spacecraft.

In addition to the $X(t)$, $Y(t)$, and $Z(t)$ signals, the recent paper by Armstrong, Estabrook, and Tinto [6] has identified several other combinations of the six one-way signals from the three arms of the interferometer that are also free of laser phase noise.  One such combination\footnote{$^6$}
  {This signal is denoted $\alpha(t)$ in reference [6], 
  but we would like to use Greek letters for the signals 
  after they are cleaned of clock-jitter, so we use $A(t)$ 
  for the laser-phase-noise-free signal given in \eq 21 and 
  reserve $\alpha(t)$ to represent its clock-jitter-cancelled 
  counterpart.}
 is
$$ A(t)=y_{31}(t)-y_{21}(t)+y_{23}(t-L_{13})-y_{32}(t-L_{12})
       +y_{12}(t-L_{13}-L_{23})-y_{13}(t-L_{12}-L_{23}) \eqno(21) $$
Other laser-noise-free signals may be found by permuting indices $1\rightarrow2\rightarrow3$ to give $B(t)$ and then $C(t)$.  As before, it is not the $y_{ij}$ that are observable, but the $s_{ij}$.  Using \eq 13a for the $s_{ij}$, \eq 21 becomes
$$ \eqalign{
A(t)=a_{21}&q_1(t)-a_{12}q_2(t-L_{13}-L_{23})+a_{13}q_3(t-L_{12}-L_{23})\cr
& -a_{31}q_1(t)+a_{32}q_2(t-L_{12})-a_{23}q_3(t-L_{13})+{\rm signal}+{\rm noise}
}\eqno(22) $$
An important difference between \eq 22 and \eq 17 may be noted.  In \eq 17, each $a_{ij}$ multiplies a difference between two $q_j$ terms taken at different times.  This is the property that made it possible to find a linear combination of the $r_{ij}$ (\eq 19) that reproduced the difference and corrected for it.  In \eq 22, however, each $a_{ij}$ multiplies a single $q_j$ term, and there is no way, short of the frequency-domain method of Reference [7], to determine a single $q_j$ time series by itself.  Therefore, there do not exist simple time-domain combinations of signals that will eliminate clock jitter from $A(t)$, $B(t)$, and $C(t)$.

Nevertheless, let us consider the following combination of $A(t)$ and $r_{ij}$ signals:
$$ \alpha(t)\equiv A(t)+a_{12}r_{23}(t-L_{13})-a_{13}r_{32}(t-L_{12})
    +(a_{12}+a_{23})r_{31}(t)-(a_{13}+a_{32})r_{21}(t)  \eqno (23a) $$
along with its permuted counterparts
$$ \beta(t)\equiv B(t)+a_{23}r_{31}(t-L_{12})-a_{21}r_{13}(t-L_{23})
    +(a_{23}+a_{31})r_{12}(t)-(a_{21}+a_{13})r_{32}(t)  \eqno (23b) $$
and 
$$ \gamma(t)\equiv C(t)+a_{31}r_{12}(t-L_{23})-a_{32}r_{21}(t-L_{13})
    +(a_{31}+a_{12})r_{23}(t)-(a_{32}+a_{21})r_{13}(t)  \eqno (23c) $$
When \eqs 19 and 22 are used in \eq 23a, we find
$$ \alpha(t)=-(a_{12}-a_{21}+a_{23}-a_{32}+a_{31}-a_{13})q_1(t)
    +{\rm signal}+{\rm noise} \eqno(24) $$
Now let us remember that the $a_{ij}$ are determined by 
$a_{ij}\sim(\nu_i-\nu_j-V_{ij}\nu_i)/f_j$.  When the $f_j$ clock frequencies are independently set on each spacecraft, there is no reason they should have any particular relationship to each other, and there is therefore no reason for the combination of $a_{ij}$ in the coefficient of $q_1(t)$ in \eq 24 to be particularly small.  On the other hand, a careful choice of these frequencies might accomplish just that.  One such choice is to make each $f_i$ proportional to the laser frequency on that spacecraft.  In fact, this condition is automatically satisfied when the local clock is formed by beating together two lasers that are phase-locked to nearby modes of the same cavity.  Whether the proportionality is achieved via two lasers or simply by tuning each USO to a frequency $f_i= \lambda\nu_i$ where $\lambda$ is some constant, the coefficient of $q_1(t)$ in \eq 24 becomes
$$ \eqalign{
a_{12}-a_{21}+a_{23}-a_{32}+a_{31}-a_{13}=
&\big[(1-V_{23})\nu_1(\nu_2+\nu_3)(\nu_3-\nu_2)\cr
&\quad+(1-V_{13})\nu_2(\nu_1+\nu_3)(\nu_1-\nu_3)\cr
&\quad\quad
+(1-V_{12})\nu_3(\nu_1+\nu_2)(\nu_2-\nu_1)\big]\big/(\lambda\nu_1\nu_2\nu_3)
}\eqno(25) $$
The terms containing $V_{ij}$ will each be of order 
$V_{ij}(\nu_i-\nu_j)/f_i\sim V_{ij}a_{ij}$, which, assuming typical values $a_{ij}\sim0.1$ and $V_{ij}\sim10^{-7}$, will make contributions of order 
$10^{-8}$.  The terms without $V_{ij}$ add up to 
$(\nu_2-\nu_1)(\nu_3-\nu_2)(\nu_3-\nu_1)/(\lambda\nu_1\nu_2\nu_3)$, which is small of order $10^{-13}$.  Thus, the $\alpha(t)$ given by \eq 23, while not eliminating the $q_i$ terms of \eq 22 identically, can nevertheless reduce them to a level where they are negligibly small.  Similar derivations and similar conclusions hold for the $\beta(t)$ and $\gamma(t)$ given in \eqs (23b) and (23c).

\vfill
\eject

\centerline{{\bf III.  Discussion and Applications}}

\vs
\n {\bf A.  Time Resolution Requirements}
\ts
\nobreak
The algorithms for producing laser phase noise cancellation (\eqs 14) and those for cancelling clock jitter (\eqs 20) require that the observed signals be available at exactly the right times so that they may be properly combined.  There are two parts to this requirement.  First, the light times $L_{ij}$ must be known with sufficient accuracy and, second, the signals must be available with sufficient time resolution for the signal with the correct time offset to be found.  The light times will only be known to some finite measurement accuracy and the $s_{ij}$ and $r_{ij}$ signals will only be sampled at some finite time resolution.  In this section, we will calculate what the requirements for the light-time measurement and signal time resolution will be.  The following two sections will discuss how the time resolution may be acheived without excessive data rate requirements for the spacecraft telemetry systems and how the light time may be measured to the required accuracy.

Let us assume that an error $\delta L_{mn}$ is made in our knowledge of the $mn^{\rm th}$ arm of the interferometer, or, equivalently, that one of the signals is not known at exactly $t-L_{mn}$, but only at $t-L_{mn}-\delta L_{mn}$.  Then a portion of the $p_j(t)$ laser phase noise (\eq 8) will remain uncorrected.  To see how this arises, we write each of the $p_j(t)$ as a Taylor series, expanding \eq 13a at time $t-L_{mn}$ to read
$$ s_{ij}(t-L_{mn})=p_i(t-L_{ij}-L_{mn})+\dot p_i(t-L_{ij}-L_{mn})\delta L_{mn}
-p_j(t-L_{mn})-\dot p_j(t-L_{mn})\delta L_{mn} \eqno(26)  $$
where we have dropped all contributions to $s_{ij}$ except for the $p_j$ terms.  It is the laser phase noise that dominates, so if we have sufficient light-time knowledge and time resolution to satisfy the laser phase noise cancellation requirement we will automatically have sufficient for the clock jitter cancellation.  When \eq 26 is used for all the $s_{ij}$ terms in \eq 16, $X(t)$ becomes
$$ \eqalign{
X(t)=(\delta L_{12}-\delta L_{13})&\dot p_1(t-2L_{12}-2L_{13})
+\delta L_{12}\left[\dot p_2(t-L_{12}-2L_{13})-\dot p_2(t-L_{12})\right]\cr
&-\delta L_{13}\left[\dot p_3(t-L_{13}-2L_{12})-\dot p_3(t-L_{13})\right]
}\eqno(27)  $$
If the measurements of $L_{12}$ and $L_{13}$ are independent and of order $\delta L$, then the relationship between the spectral density of laser phase noise in $X(t)$ and the spectral density of $\dot p_i$ is 
$$ S_X=\left[2+4\sin^2(2\omega L_{13})+4\sin^2(2\omega L_{12})\right]
(\delta L)^2 S_{\Delta\nu} \eqno(28) $$
where $S_{\Delta\nu}=S_{\dot p}$ is the spectrum of frequency fluctuations.  The time-domain differences of $\dot p_j$ in \eq 27 produced the $\sin^2(2\omega L)$ terms as transfer functions in \eq 28.  If we average \eq 28 over all frequencies $\omega$, we find
$$ \bar S_X=7.66(\delta L)^2S_{\Delta\nu}\eqno(29) $$
giving a requirement on $\Delta L$ of
$$ \delta L=
0.36\, {\sqrt{\bar S_X}\over\sqrt{S_{\Delta\nu}}}=3.6\times10^{-6}\,{\rm s}
\eqno(30) $$
where, in the evaluation of \eq 30, we have assumed the usual phase noise requirement of $10^{-5}$\s cycles\prh and assumed that the spectrum of laser frequency stability observed in the laboratory, 
$S_{\Delta\nu}(f)\sim1{\rm Hz}^2\ {\rm Hz}^{-1}$ at a frequency of 1\s Hz, can be continued to lower frequency when the Fabry-Perot cavity has the thermal stability it is expected to have in space.  If this is indeed the case, a time-resolution of $1\mu$s would reduce the residual laser phase noise to a negligible level.  If this stability is not acheivable in space, a finer time resolution would be needed for sampling the $s_{ij}$ and $r_{ij}$ signals, and a more precise knowledge of the light time between spacecraft would be required.

\vs
\goodbreak
\n {\bf B.  Signal Requirements}
\nobreak
\ts
\nobreak
As was discussed in the previous section, the signals must be sampled on board each spacecraft with a $1\,\mu$s or better resolution.  However, \eqs 14 and 21 show that the combinations needed to eliminate laser noise require signals from all three spacecraft ({\it i.e.}, the formation of $X(t)$ requires $s_{12}$, $s_{13}$, and $s_{21}-s_{31}$).  As we will show in Section III.D, even in the case where some of the lasers are locked to the incoming signals, signals from multiple spacecraft are needed in order to form the complete unequal-arm interferometer combinations.  Thus, the signals read out on board each spacecraft will have to be telemetered to the ground or to other spacecraft in order to create the required combinations.  The straightforward communication of the time series for $s_{ij}$ and $r_{ij}$ with $\mu$s time resolution, and with enough bits to define the phase to the required accuracy, would require a data rate on the order of 100\s Mb/s.  This is prohibitively high for the space missions that have been proposed.  Fortunately, it is also unnecessary, as we shall now show.

In order to detect gravitational waves in the LF frequency band, a sample time of $\Delta t=1\,$s is all that is required in the $X(t)$ and other signals.  The only reason for using a higher sample rate than this would be to provide the higher resolution necessary to correctly create $X(t)$ and the other signals by combining data from the various spacecraft with the proper time offsets.  However, one-second samples of $X(t)$ can be generated by averaging a higher resolution time series.  If we average \eq 16 over the desired sample time, $\Delta t$, we find:
$$ \eqalign{
X_n\equiv&{1\over\Delta t}\int_{t_n}^{t_n+\Delta t}X(t)dt\cr
=&{1\over\Delta t}\Biggl\{\int_{t_n}^{t_n+\Delta t}s_{12}(t-L_{12}-2L_{13})dt
-\int_{t_n}^{t_n+\Delta t}s_{13}(t-L_{13}-2L_{12})dt\cr
&\quad+\int_{t_n}^{t_n+\Delta t}s_{21}(t-2L_{13})dt
-\int_{t_n}^{t_n+\Delta t}s_{31}(t-2L_{12})dt
+\int_{t_n}^{t_n+\Delta t}s_{13}(t-L_{13})dt\cr
&\quad\quad-\int_{t_n}^{t_n+\Delta t}s_{12}(t-L_{12})dt
+\int_{t_n}^{t_n+\Delta t}s_{31}(t)dt
-\int_{t_n}^{t_n+\Delta t}s_{21}(t)dt\Biggr\}
}\eqno(31)  $$
where $X_n$ is the $n^{\rm th}$ sample, taken at time $t_n=n\Delta t$.  Therefore, if the $s_{ij}$ signals are available on each spacecraft with $\mu$s resolution, all that is required in order to allow the $X(t)$ combination to be formed is the integral of each of the eight terms on the right-hand side over the $\Delta t$ interval.  In fact, all that is required, for example, from S/C 2 is the combination
$$ {1\over\Delta t}\int_{t_n}^{t_n+\Delta t}\left[s_{12}(t-L_{12}-2L_{13})
-s_{12}(t-L_{12})\right]dt \eqno(32) $$
Consider, for example, a case with light times $L_{12}=15.7763391\,$s and $L_{13}=15.8227142\,$s.  At a time $t_{7184}=7184\,$s, S/C 2 would need to have available the average of $s_{12}$ from $t=7136.578033\,$s ({\it i.e.}, 
$t_n-L_{12}-2L_{13}$ with 1$\,\mu$s resolution) to 7137.578033\s s 
($t_n-L_{12}-2L_{13}+\Delta t$).  The average of $s_{12}$ from 7168.223661 ($t_n-L_{12}$) to 7169.223661 ($t_n-L_{12}+\Delta t$) would also be needed.  The difference between these two 1-second averages is the signal required from S/C 2 in order to form $X(t)$.

If we consider the total data requirements from all spacecraft and for the complete set of variables, $\chi(t)$, $\psi(t)$, and $\zeta(t)$, the following data requirements table may be formed:

\ts
\line{\hfill
\vtop{\tabskip=0pt \offinterlineskip
\halign{
\vrule width.8pt#\tabskip=1em plus 2em
&\hfil # \hfil   &\vrule width.8pt#
&\hfil # \hfil   &\vrule width.8pt#
&\hfil # \hfil   &\vrule width.8pt#
&\hfil # \hfil   &\vrule width.8pt#
&\hfil # \hfil   &\vrule width.8pt#     
\tabskip=0pt\cr
\noalign{\hrule height.8pt}
height5pt&\omit&&\omit&\omit&\omit&\omit&\omit&\omit&\omit&\cr
& signals && \omit & \omit & times & \omit & \omit &\omit &&\cr
height5pt&\omit&&\omit&\omit&\omit&\omit&\omit&\omit&\omit&\cr
\noalign{\hrule height.8pt}
height5pt&\omit&&\omit&&\omit&&\omit&&\omit&\cr
& $s_{21}$ && $t-L_{12}-2L_{23}$ && $t-L_{12}$ && $t-2L_{13}$ && $t$ &\cr
height5pt&\omit&&\omit&&\omit&&\omit&&\omit&\cr
& $r_{21}$ &&                    && $t-L_{12}$ && $t-2L_{13}$ && $t$ &\cr
height5pt&\omit&&\omit&&\omit&&\omit&&\omit&\cr
& $s_{31}$ && $t-L_{13}-2L_{23}$ && $t-L_{13}$ && $t-2L_{12}$ && $t$ &\cr
height5pt&\omit&&\omit&&\omit&&\omit&&\omit&\cr
& $r_{31}$ &&                    && $t-L_{13}$ && $t-2L_{12}$ && $t$ &\cr
height5pt&\omit&&\omit&&\omit&&\omit&&\omit&\cr
\noalign{\hrule height.8pt}
height5pt&\omit&&\omit&&\omit&&\omit&&\omit&\cr
& $s_{12}$ && $t-L_{12}-2L_{13}$ && $t-L_{12}$ && $t-2L_{23}$ && $t$ &\cr
height5pt&\omit&&\omit&&\omit&&\omit&&\omit&\cr
& $r_{12}$ &&                    && $t-L_{12}$ && $t-2L_{23}$ && $t$ &\cr
height5pt&\omit&&\omit&&\omit&&\omit&&\omit&\cr
& $s_{32}$ && $t-L_{23}-2L_{13}$ && $t-L_{23}$ && $t-2L_{12}$ && $t$ &\cr
height5pt&\omit&&\omit&&\omit&&\omit&&\omit&\cr
& $r_{32}$ &&                    && $t-L_{23}$ && $t-2L_{12}$ && $t$ &\cr
height5pt&\omit&&\omit&&\omit&&\omit&&\omit&\cr
\noalign{\hrule height.8pt}
height5pt&\omit&&\omit&&\omit&&\omit&&\omit&\cr
& $s_{13}$ && $t-L_{13}-2L_{12}$ && $t-L_{13}$ && $t-2L_{23}$ && $t$ &\cr
height5pt&\omit&&\omit&&\omit&&\omit&&\omit&\cr
& $r_{13}$ &&                    && $t-L_{13}$ && $t-2L_{23}$ && $t$ &\cr
height5pt&\omit&&\omit&&\omit&&\omit&&\omit&\cr
& $s_{23}$ && $t-L_{23}-2L_{12}$ && $t-L_{23}$ && $t-2L_{13}$ && $t$ &\cr
height5pt&\omit&&\omit&&\omit&&\omit&&\omit&\cr
& $r_{23}$ &&                    && $t-L_{23}$ && $t-2L_{13}$ && $t$ &\cr
height5pt&\omit&&\omit&&\omit&&\omit&&\omit&\cr
\noalign{\hrule height.8pt}
}}\hfill}
\n Table 1.  Signal requirements for the set $\chi(t)$, $\psi(t)$, and $\zeta(t)$.  The times listed for each signal are those at which the signal must be accumulated on board each spacecraft.

\vs

As may be seen, each spacecraft is required to produce four one-second averages of each incoming $s_{ij}$ signal and three of each $r_{ij}$ signal, each with its particular time offset.  Since the $r_{ij}$ signals are required at the same times as their respective $s_{ij}$ signals, all that is required to produce them is to generate the $s'_{ij}$ at the three times given above for the $r_{ij}$ and to simply subtract to get $r_{ij}=s_{ij}-s'_{ij}$ as in \eq 18.  The fourteen pieces of data at each spacecraft are then combined into three signals that are telemetered to other spacecraft or to the ground in order for the $\chi(t)$, $\psi(t)$, and $\zeta(t)$ set of signals to be formed.  The data required from each spacecraft are 
$$ {\rm S/C1:}\quad\quad\cases{D_{1\chi}=
s_{21}(t-2L_{13})-s_{21}(t)-s_{31}(t-2L_{12})+s_{31}(t)\cr
\hfil-a_{21}r_{21}(t-2L_{13})+(a_{12}+a_{13}+a_{31})r_{21}(t)\cr
D_{1\psi}=
s_{21}(t-L_{12})-s_{21}(t-L_{12}-2L_{23})-(a_{23}+a_{32})r_{21}(t-L_{12})\cr
D_{1\zeta}=
s_{31}(t-L_{13})-s_{31}(t-L_{13}-2L_{23})-(a_{23}+a_{32})r_{31}(t-L_{13})\cr
}\eqno(33a)$$
$$ {\rm S/C2:}\quad\quad\cases{D_{2\psi}=
s_{32}(t-2L_{12})-s_{32}(t)-s_{12}(t-2L_{23})+s_{12}(t)\cr
\hfil-a_{32}r_{32}(t-2L_{12})+(a_{23}+a_{12}+a_{21})r_{32}(t)\cr
D_{2\zeta}=
s_{32}(t-L_{23})-s_{32}(t-L_{23}-2L_{13})-(a_{13}+a_{31})r_{32}(t-L_{23})\cr
D_{2\chi}=
s_{12}(t-L_{12})-s_{12}(t-L_{12}-2L_{13})-(a_{13}+a_{31})r_{12}(t-L_{12})\cr
}\eqno(33b)$$
$$ {\rm S/C3:}\quad\quad\cases{D_{3\zeta}=
s_{13}(t-2L_{23})-s_{13}(t)-s_{23}(t-2L_{13})+s_{23}(t)\cr
\hfil-a_{13}r_{13}(t-2L_{23})+(a_{31}+a_{23}+a_{32})r_{13}(t)\cr
D_{3\chi}=
s_{13}(t-L_{13})-s_{13}(t-L_{13}-2L_{12})-(a_{12}+a_{21})r_{13}(t-L_{13})\cr
D_{3\psi}=
s_{23}(t-L_{23})-s_{23}(t-L_{23}-2L_{12})-(a_{12}+a_{21})r_{23}(t-L_{23})\cr
}\eqno(33c)$$
It should be noted that all six values of the $a_{ij}$ must be known by each spacecraft in order to form these combinations on board.  It can easily be verified that $\chi(t)$, $\psi(t)$, and $\zeta(t)$ are formed by adding together just these nine pieces of data, so this set of signals represents the data rate requirement for this type of mission.

The set of data needed to form $\alpha(t)$, $\beta(t)$, and $\gamma(t)$ is summarized in Table 2.

\ts
\line{\hfill
\vtop{\tabskip=0pt \offinterlineskip
\halign{
\vrule width.8pt#\tabskip=1em plus 2em
&\hfil # \hfil   &\vrule width.8pt#
&\hfil # \hfil   &\vrule width.8pt#
&\hfil # \hfil   &\vrule width.8pt#
&\hfil # \hfil   &\vrule width.8pt#
\tabskip=0pt\cr
\noalign{\hrule height.8pt}
height5pt&\omit&&\omit&\omit&\omit&\omit&\omit&\cr
& signals && \omit & \omit & times & \omit &&\cr
height5pt&\omit&&\omit&\omit&\omit&\omit&\omit&\cr
\noalign{\hrule height.8pt}
height5pt&\omit&&\omit&&\omit&&\omit&\cr
& $s_{21}$ && $t-L_{13}-L_{23}$ && $t-L_{13}$ && $t$ &\cr
height5pt&\omit&&\omit&&\omit&&\omit&\cr
& $r_{21}$ &&                   && $t-L_{13}$ && $t$ &\cr
height5pt&\omit&&\omit&&\omit&&\omit&\cr
& $s_{31}$ && $t-L_{12}-L_{23}$ && $t-L_{12}$ && $t$ &\cr
height5pt&\omit&&\omit&&\omit&&\omit&\cr
& $r_{31}$ &&                   && $t-L_{12}$ && $t$ &\cr
height5pt&\omit&&\omit&&\omit&&\omit&\cr
\noalign{\hrule height.8pt}
height5pt&\omit&&\omit&&\omit&&\omit&\cr
& $s_{12}$ && $t-L_{13}-L_{23}$ && $t-L_{23}$ && $t$ &\cr
height5pt&\omit&&\omit&&\omit&&\omit&\cr
& $r_{12}$ &&                   && $t-L_{23}$ && $t$ &\cr
height5pt&\omit&&\omit&&\omit&&\omit&\cr
& $s_{32}$ && $t-L_{12}-L_{13}$ && $t-L_{12}$ && $t$ &\cr
height5pt&\omit&&\omit&&\omit&&\omit&\cr
& $r_{32}$ &&                   && $t-L_{12}$ && $t$ &\cr
height5pt&\omit&&\omit&&\omit&&\omit&\cr
\noalign{\hrule height.8pt}
height5pt&\omit&&\omit&&\omit&&\omit&\cr
& $s_{13}$ && $t-L_{12}-L_{23}$ && $t-L_{23}$ && $t$ &\cr
height5pt&\omit&&\omit&&\omit&&\omit&\cr
& $r_{13}$ &&                   && $t-L_{23}$ && $t$ &\cr
height5pt&\omit&&\omit&&\omit&&\omit&\cr
& $s_{23}$ && $t-L_{12}-L_{13}$ && $t-L_{13}$ && $t$ &\cr
height5pt&\omit&&\omit&&\omit&&\omit&\cr
& $r_{23}$ &&                   && $t-L_{13}$ && $t$ &\cr
height5pt&\omit&&\omit&&\omit&&\omit&\cr
\noalign{\hrule height.8pt}
}}\hfill}
\n Table 2.  Signal requirements for the set $\alpha(t)$, $\beta(t)$, and $\gamma(t)$.  The times listed for each signal are those at which the signal must be accumulated on board each spacecraft.

\ts

\n In the case of the $\alpha(t)$, $\beta(t)$, and $\gamma(t)$ combinations, there are fewer time offset signals that must be accumulated on each spacecraft (three $s_{ij}$ instead of four and two $r_{ij}$ instead of three), but the times are not all the same as those in Table 1.  Thus, if both sets, $\chi(t)$, $\psi(t)$, and $\zeta(t)$ and $\alpha(t)$, $\beta(t)$, and $\gamma(t)$, are desired, then more data must be accumulated in the laser receivers (six $s_{ij}$ and four $r_{ij}$).  The data combinations needed to form the $\alpha(t)$, $\beta(t)$, $\gamma(t)$ set are
$$ {\rm S/C1:}\quad\quad\cases{D_{1\alpha}=
s_{31}(t)-s_{21}(t)+(a_{21}+a_{23})r_{31}(t)-(a_{13}+a_{32})r_{21}(t)\cr
D_{1\beta}=
s_{31}(t-L_{12})-s_{21}(t-L_{13}-L_{23})+a_{23}r_{31}(t-L_{12})\cr
D_{1\gamma}=
s_{31}(t-L_{12}-L_{23})-s_{21}(t-L_{13})-a_{32}r_{21}(t-L_{13})\cr
}\eqno(34a)$$
$$ {\rm S/C2:}\quad\quad\cases{D_{2\beta}=
s_{12}(t)-s_{32}(t)+(a_{32}+a_{31})r_{12}(t)-(a_{13}+a_{21})r_{32}(t)\cr
D_{2\gamma}=
s_{12}(t-L_{23})-s_{32}(t-L_{13}-L_{12})+a_{31}r_{12}(t-L_{23})\cr
D_{2\alpha}=
s_{12}(t-L_{13}-L_{23})-s_{32}(t-L_{12})-a_{13}r_{32}(t-L_{12})\cr
}\eqno(34b)$$
$$ {\rm S/C3:}\quad\quad\cases{D_{3\gamma}=
s_{23}(t)-s_{13}(t)+(a_{31}+a_{12})r_{23}(t)-(a_{32}+a_{21})r_{13}(t)\cr
D_{3\alpha}=
s_{23}(t-L_{13})-s_{13}(t-L_{12}-L_{23})+a_{12}r_{23}(t-L_{13})\cr
D_{3\beta}=
s_{23}(t-L_{12}-L_{13})-s_{13}(t-L_{23})-a_{21}r_{13}(t-L_{23})\cr
}\eqno(34c)$$
As was the case for $\chi(t)$, $\psi(t)$, and $\zeta(t)$, each spacecraft must know all of the $a_{ij}$ from the other spacecraft before the $\alpha(t)$, $\beta(t)$, $\gamma(t)$ combinations can be formed.  The data requirement for $\alpha(t)$, $\beta(t)$, $\gamma(t)$ is again nine pieces of data, three from each of three spacecraft.  The signal combinations in \eq 34 are independent of those in \eq 33, so, if both the $\chi(t)$, $\psi(t)$, and $\zeta(t)$ and the $\alpha(t)$, $\beta(t)$, $\gamma(t)$ data sets are desired simultaneously, the data rata requirement is doubled.

Assuming that each data point requires, say, 64 bits to accurately represent the phase in microcycles, the data required from the spacecraft has been reduced, as a result of the considerations in this section, from 100\s Mb/s to $9\times64 = 576\,$b/s.  The laser phases will be sampled with microsecond accuracy and then averaged to form the fourteen one-second phase measurements given in Tables 1 and 2.  These are then added together in the proper ratios to produce the signal combinations of \eq 33 or \eq 34, and this is all that must be telemetered to the ground.  There remains, however, one important use for the high rate data.  This is the subject of the next section.

\vs
\n {\bf C.  Measuring the $L_{ij}$}
\ts
\nobreak
Let us consider the cross-correlation function of $s_{21}$ and $s_{12}$ over a record of length $T$.
$$ C_{12}(\tau)={1\over T}\int_{-T/2}^{T/2}s_{21}(t-\tau)s_{12}(t)dt 
\eqno(35) $$
When the $s_{ij}$ are expanded, keeping only the dominant laser phase noise terms, \eq 35 becomes
$$ \eqalign{
C_{12}(\tau)&=
{1\over T}\int_{-T/2}^{T/2}\left[p_2(t-L_{12}-\tau)-p_1(t-\tau)\right]
                          \left[p_1(t-L_{12})-p_2(t)\right]dt\cr
&=-{1\over T}\int_{-T/2}^{T/2}p_2(t-L_{12}-\tau)p_2(t)dt
-{1\over T}\int_{-T/2}^{T/2}p_1(t-\tau)p_1(t-L_{12})dt
}\eqno(36) $$
where the the $p_1p_2$ terms have been dropped since the statistical independence of $p_1$ and $p_2$ will make then negligibly small.  The two remaining integrals in \eq 36 are just the autocorrelation functions of $p_1$ and $p_2$ with time offsets,
$$ C_{12}(\tau)=-A_{p2}(\tau+L_{12})-A_{p1}(\tau-L_{12}) \eqno(37) $$
Thus, the cross-correlation function will have peaks at $+L_{12}$ and $-L_{12}$, corresponding to the zero-lag peaks in the autocorrelation functions of $p_1$ and $p_2$.  Examination of the cross-correlation function and identification of these peaks will therefore provide a measure of the light time $L_{12}$.  

The measurement of the $L_{ij}$ will not need to be performed on a continuing basis, since the observed Doppler rate in $s_{ij}$ will give the change in range between spacecraft.  The measurement will therefore only be needed once initially and again whenever a check of the integrated range is desired.  For each range measurement, the $s_{ij}$ may be sampled at a raw rate of one every microsecond and the raw data from several seconds of measurement of $s_{ij}$ and $s_{ji}$ can be telemetered to the ground at a low data rate over a long period of time.  Once both signals are available, the identification of $L_{ij}$ will consist simply in forming $C_{ij}$ and finding the unique value of $\tau$ where the two maxima of the cross-correlation function occur.  If the data are averaged over several sample times, in order to reduce the data required for forming $C_{ij}$, then interpolation between the two maxima on either side of $\tau=\pm L_{ij}$ can determine the correct value.

One practical problem in the procedure we have just described is that the solid-state lasers we have assumed for the space missions are, in fact, very low in phase noise at a sample time of 1\s$\mu$s.  The products of the $p_i$ laser phase noises risk being swallowed up in the $n_{ij}$ shot noises in the phase readout system.  There are two things that may be done to get around this problem.  First, since there is always a phase modulator in the main laser beam that is used for the phase locking to the Fabry-Perot cavity, that phase modulator could be used to impose a 1\s MHz pseudo-random phase noise code on the laser signal.  The resulting phase shifts could be as large as desired and could be much larger than the shot noise floor.  Alternatively, the process of averaging the raw data rate over several samples will give a laser phase noise at what is effectively a longer sample time.  The noise at this sample time will be greater than that at the raw rate of 1\s$\mu$s since lasers typically exhibit a $1/f$ noise spectrum in this frequency range.  This noise may be great enough by itself to be detectable above the shot noise, and the true value of $\tau$ could be determined by interpolating the correlation function.  In Fig.\ 4, a simulation of this general technique is shown for a 1\s ms (instead of a 1\s$\mu$s) time resolution.  A total of $10^5$ points of pseudo-random $p_i(t)$ phase noise data were generated with a fundamental time resolution of 0.1\s ms.  The data were then lagged by the proper amounts to represent $s_{12}$ and $s_{21}$ to 0.1\s ms resolution.  The $s_{ij}$ were each averaged over 10-point bins to form a total of 10 seconds of 1-ms-resolution signal.  A sample of the time series of $s_{12}$ is shown in Fig.\ 4a.  Finally, the cross-correlation function $C_{12}(\tau)$ was formed (Fig.\ 4b), and an interpolated estimate of $L_{12}$ was made, as given in the figure.  A similar set of graphs is shown in Fig.\ 5 for assumed $1/f$ noise.  In each case, with only $10^4$ data points in the cross-correlation, the interpolation has determined the correct value of $L_{12}$ (133.4\s ms) to about 0.4\% of the 1\s ms resolution.

\vs
\n {\bf D.  Lasers Phase-Locked to the Incoming Signals}
\ts
\nobreak
The implementation of the space interferometer that most resembles the laboratory Michelson interferometer is that where two of the spacecraft act like active mirrors, transponding without any change of phase whatever signal they receive.  For example, S/C 1 in Fig. 1 might contain the master laser and the lasers on S/C 2 and S/C 3 would be phase locked to the signal they received from S/C 1.  As we have shown in \eq 2 and \eq 5, the forms of the Michelson signal $z_1(t)$ and the interferometer signal $X(t)$ are particularly simple in this case and may be completely formed with signals that are available on S/C 1.  This simple form may also be found, beginning with the fundamental formula for one-way arms (\eq 14a) by setting $y_{13}$ and $y_{12}$ to zero, reflecting the fact that the phase-locked lasers on S/C 2 and S/C 3 will be locked in such a way that this condition is always met.  However, when these two quantities are set to zero in the formula for $Y(t)$ and $Z(t)$, the results are
$$ \eqalign{
Y(t)=y_{23}&(t-L_{23}-2L_{12})-y_{21}(t-L_{12}-2L_{23})+y_{32}(t-2L_{12})\cr
&+y_{21}(t-L_{12})-y_{23}(t-L_{23})-y_{32}(t)
}\eqno(38a) $$
and
$$ \eqalign{
Z(t)=y_{31}&(t-L_{13}-2L_{23})-y_{32}(t-L_{23}-2L_{13})-y_{23}(t-2L_{13})\cr
&+y_{32}(t-L_{23})-y_{31}(t-L_{13})+y_{23}(t)
}\eqno(38b) $$
One point to be noticed here is that $Y(t)$ and $Z(t)$ cannot be written by permuting indices in \eq 6, since there is a fundamental difference between an independent central laser sending out two beams that are reflected and returned and a central laser that is itself locked to an incoming beam whose phase is independent.  Thus, even when the end lasers are phase-locked, the only way to form combinations beyond $X(t)$ that are laser-phase-noise free is to use the formulas in \eq 38.  As is explicitly shown in \eq 38a, the formation of $Y(t)$ requires a combination of signals from S/C 3, 
$y_{23}(t-L_{23}-2L_{12})-y_{23}(t-L_{23})$, and from S/C 1, 
$y_{21}(t-L_{12}-2_{L23})-y_{21}(t-L_{12})$, in addition to the signal, $y_{32}(t-2L_{12})-y_{32}(t)$, from S/C 2.  Different combinations are required, again from all three spacecraft, in order to form $Z(t)$.

When the lasers in S/C 2 and S/C 3 are locked to the incoming phase of the lasers from S/C 1, the correct formulas for the generation of $\chi(t)$, $\psi(t)$, $\zeta(t)$, $\alpha(t)$, $\beta(t)$, and $\gamma(t)$ are found by setting $y_{13}$ and $y_{12}$ identically to zero at all times in \eqs 14 and in \eq 21 (with its permuted extensions).  To see the data requirements from each spacecraft, the same $s_{13}$ and $s_{12}$ variables should be set to zero in \eqs 33 and 34.  As may be seen in \eqs 33, the data rate from each spacecraft is the same, with or without the end lasers locked to their incoming signals.  Only if the clocks on board each end spacecraft were also phase locked to the incoming difference between $s_{1i}$ and $s'_{1i}$ (so that $r_{12}=r_{13}=0$), would the two signals $D_{2\chi}$ and $D_{3\chi}$ in \eq 33 not need to be telemetered from the two end spacecraft.  And, in the case of $\alpha(t)$, $\beta(t)$, and $\gamma(t)$ there can be no reduction in data rate at all.

To summarize, there is nothing wrong with locking the end lasers in the way that has been proposed.  Such locking will probably be required if there is a desire to reduce the LO frequency to the minimal $\sim10$\s MHz that is obtained when the only frequency offsets are the unavoidable Doppler frequency shifts.  However, this approach provides essentially no simplification of the data taking nor reduction of the data rate requirements for the mission.

\vfill
\eject

\centerline{{\bf IV.  Conclusions}}

\vs

As a result of the algorithms we have presented, we find that it is possible to completely eliminate the clock jitter noise at all frequencies for the case of the $X(t)$, $Y(t)$, and $Z(t)$ variables by use of the data combinations in \eqs 20.  For the case of the $A(t)$, $B(t)$, and $C(t)$ variables, it is possible to reduce the clock jitter noise to a negligible level by use of \eqs 23 and by a judicious choice of the clock frequencies on board each spacecraft.  We have also investigated the timing precision and resolution required for creation of the laser phase noise elimination and for the clock jitter elimination (\eq 30) and have suggested a method for measuring the time-of-flight of the laser signals to this accuracy.  Finally, we have specified how the data rate requirements for the mission can be minimized by use of appropriate simultaneous time-offset averages on each spacecraft.  The data that must be generated by the laser receivers is given in Table 1 for the $\chi(t)$, $\psi(t)$, and $\zeta(t)$ variables and in Table 2 for the $\alpha(t)$, $\beta(t)$, and $\gamma(t)$ set.  The data that are to be communicated to the ground in order to form the laser-noise free and clock-jitter free signals are then given by \eqs 33 for $\chi(t)$, $\psi(t)$, and $\zeta(t)$ and by \eqs 34 for $\alpha(t)$, $\beta(t)$, and $\gamma(t)$.  From these requirements, we derive a minimum data accumulation rate of 576\s b/s for the set of all three spacecraft.  In the case where two lasers are phase-locked to the incoming signals from one master laser, the data rate for the $\chi(t)$, $\psi(t)$, and $\zeta(t)$ variables can be reduced to 488 b/s (if the local spacecraft clocks are likewise phase locked to the master spacecraft clock), but there is no reduction possible at all for the $\alpha(t)$, $\beta(t)$, and $\gamma(t)$ set of variables.

\vfill
\eject

\n {\bf References:}

\n 1.  R.W. Hellings, Contemp. Phys. {\bf37} 457-469 (1996).

\n 2.  P. Bender et al., ``LISA Pre-Phase A Report (second edition)'' (1998).

\n 3.  R.W. Hellings et al., ``Orbiting Medium Explorer for Gravitational Astrophysics (OMEGA)'', proposal to NASA Medium Explorer program (unpublished).

\n 4.  G. Giampieri, R. Hellings, M. Tinto, and J.E. Faller, Optics Comm. {\bf123} 669-78 (1996).

\n 5.  M. Tinto and J.W. Armstrong, Phys. Rev. {\bf D59} 102003 (1999).

\n 6.  J.W. Armstrong, F.B. Estabrook, and M. Tinto, Ap. J. {\bf527} 814-826 (2000).

\n 7.  R. Hellings, G. Giampieri, L. Maleki, M. Tinto, K. Danzmann, J. Hough, and D. Robertson, Optics Comm. {\bf124} 313-20 (1996).

\n 8. R.W. Hellings, in ``Proceedings of the Third LISA Symposium, Golm, Germany'' (2001).

\n 9. S.L. Larson, W.A. Hiscock, and R.W. Hellings, Phys Rev. {\bf D62} 062001 (2000).

\n 10. F.D. Cady and R.W. Hellings, manuscript in preparation (2001).

\bye